\documentclass[aps,
               prl,
               amsmath,
               amssymb,
               notitlepage,
               twocolumn,
               superscriptaddress,
               longbibliography,
               nobalancelastpage,
               nofootinbib,
               floatfix,
               10pt]{revtex4-2}

\usepackage{svg}
\usepackage{amsmath}
\usepackage{amsfonts}
\usepackage{amssymb}
\usepackage{amsthm}
\usepackage{mathrsfs}
\usepackage{natbib}
\usepackage{circuitikz}
\usepackage{MnSymbol}
\usepackage{slashed}
\usepackage{graphicx}
\usepackage{bm}
\usepackage{orcidlink}
\usepackage{mathtools}

\DeclarePairedDelimiter\floor{\lfloor}{\rfloor}

\svgsetup{inkscapelatex=false}

\date{\today}

\begin{document}

\author{Kostas Tzanavaris\,\orcidlink{0000-0003-0949-7191}}\affiliation{\AEI}\affiliation{\Leibniz}\affiliation{\Edinburgh}
\author{Latham Boyle\,\orcidlink{0000-0001-5195-7599
}}\affiliation{\Edinburgh}\affiliation{\PI}
\author{Neil Turok\,\orcidlink{0000-0002-1891-8691}}\affiliation{\Edinburgh}\affiliation{\PI}
\newcommand*{\Edinburgh}{Higgs Centre for Theoretical Physics, James Clerk Maxwell Building, Edinburgh EH9 3FD, UK}
\newcommand*{\AEI}{Max Planck Institute for Gravitational Physics (Albert Einstein Institute), D-30167 Hannover, Germany} 
\newcommand*{\PI}{Perimeter Institute for Theoretical Physics, Waterloo, Ontario N2L 2Y5, Canada}
\newcommand*{\Leibniz}{Leibniz Universit{\"a}t Hannover, D-30167 Hannover, Germany}

\title{The free boundary problem in general relativity}

\begin{abstract}

We study the action principle for space-times whose boundary is singular.  We suggest that it is natural to treat the singularity  as a {\it free} boundary, where the variation is unconstrained.  Demanding that the action is stationary under such free variations then implies certain (on-shell) boundary conditions at the singularity.   
We derive these boundary conditions for the case of Einstein gravity coupled to matter and show that, when applied to an initial spacelike singularity, they exclude Kasner-like or BKL space-times, but admit conformally regular space-times (including FLRW models) sourced by fluids satisfying $0\leq P < \rho$. 
For standard hot big bang FLRW cosmologies, 
the admissible linear (scalar, vector, tensor) perturbations satisfy reflecting boundary conditions at the bang, in agreement with large-scale cosmological observations.
\end{abstract}

\maketitle

\section{Introduction}

The action principle \cite{hawking:1973spacetime, christodoulou:2000action}, which follows from the path integral formulation of quantum field theory, asserts that the classically-allowed field configurations $\varphi_{c}$ are the stationary points of the action functional $\mathcal{S}[\varphi]$, subject to some given boundary conditions.  We usually imagine varying the fields on a specified space-time manifold ${\cal M}$ with specified boundary $\partial{\cal M}$.  The procedure is clear when we are path integrating over matter fields on a fixed background ${\cal M}$ with a fixed metric $\mathbf{g}$, and non-singular boundary $\partial{\cal M}$.  However, quantum gravity introduces a new wrinkle, since we are now supposed to path integrate over different configurations of the metric $\mathbf{g}$, satisfying specified boundary conditions on $\partial{\cal M}$; but this inevitably involves integrating over metrics which may be singular on {\it new} boundary regions in the interior of ${\cal M}$, {\it in addition} to those we originally specified.

Here we reconsider the action principle for a space-time ${\cal M}$ whose boundary $\partial{\cal M}$ includes such a singular component.  We explain why it is natural to treat the singularity as a {\it free} boundary where the off-shell variation is {\it not} required to vanish.  Demanding $\mathcal{S}$ is stationary under such variations then yields (on-shell) boundary conditions at the singularity.  We derive these boundary conditions for Einstein gravity and show that, when applied to space-times with a singular spacelike boundary, they exclude Kasner- or BKL-type space-times, but permit conformally-regular space-times (including FLRW models) sourced by fluids satisfying $0 \leq P<\rho$.  In particular, for FLRW hot big bang cosmologies, the admissible linear perturbations satisfy reflecting boundary conditions at the bang which are just those compatible with the observed large scale cosmological perturbations.

We begin with a precise formulation of the action principle in field theories, including gravity.  For space-times without boundary our formulation is equivalent to those of Hawking and Ellis \cite{hawking:1973spacetime} or Christodoulou \cite{christodoulou:2000action}.

\section{The action principle in field theory}

\noindent
\textbf{The action principle:}
Let $(\mathcal{M},\mathbf{g})$ be a four-dimensional space-time. 
Given a field $\varphi$ on $\mathcal{M}$ with a Lagrangian density $\mathcal{L}$, its classically allowed configurations are the critical points of its action $\mathcal{S}[\varphi,\mathcal{D}]:=\int_{\mathcal{D}}\mathcal{L}[\varphi]$, for every open region $\mathcal{D}\subset\mathcal{M}$ with compact closure. 

\vspace{0.5em}

\begin{figure}[ht!]
    \centering
    \includegraphics[width=0.64\linewidth]{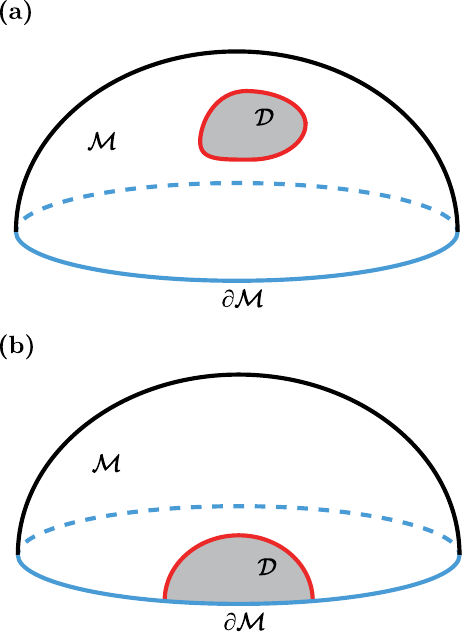}
    \caption{In both cases, the variations $\delta\varphi$ are required to vanish on the red portion of the boundary, and outside it. In case (a), the red portion coincides with the whole boundary $\partial\mathcal{D}$, so all boundary terms vanish. In case (b), however, the variations $\delta\varphi$ on $\mathcal{D}\cap\partial\mathcal{M}$ are free, as $\mathcal{D}$ is open and intersects $\partial\mathcal{M}$.}
    \label{fig:action-with-bound}
\end{figure}

\noindent
Here, 
$\varphi_c$ is called a critical point of $\mathcal{S}[\cdot,\mathcal{D}]$ iff
\begin{equation}
    \delta\mathcal{S}[\varphi_c,\mathcal{D}]:=\frac{d}{ds}\bigg\rvert_{s=0}
    \mathcal{S}[\varphi_c + s\,\delta\varphi,\mathcal{D}] = 0
    \label{eqn:action-variation-def}
\end{equation}
for every smooth variation $\delta\varphi$ supported in $\mathcal{D}$, \textit{i.e.}, $\delta\varphi(x) = 0$ whenever $x\in\mathcal{M}\setminus\mathcal{D}$. 
Taking ${\cal D}$ with compact closure ensures the variation $\delta S$ is well defined for any such $\delta\varphi$. Note that, since $\delta\varphi$ vanishes outside $\mathcal{D}$, continuity requires it to vanish at the edge of $\mathcal{D}$, when ${\cal D}$ does not intersect
the boundary of ${\cal M}$ (Fig.~\ref{fig:action-with-bound}a).  Thus we recover the usual requirement that we vary the action on an interval, while fixing the variation at the endpoints.

Now consider this action principle on
a space-time $(\mathcal{M},\mathbf{g})$ with a non-empty boundary $\partial\mathcal{M}$, and for simplicity, we assume that $\varphi$ is a scalar field. Notice that now, if $\mathcal{D}$ intersects $\partial\mathcal{M}$, continuity requires the variation $\delta\varphi$ to vanish at the edge of $\mathcal{D}$ {\it except} where $\mathcal{D}$ intersects the space-time boundary $\partial{\cal M}$ (see Fig.~\ref{fig:action-with-bound}b). 
Hence, under a variation of $\varphi$ in $\mathcal{D}$, the on-shell variation of the action is $\delta\mathcal{S}[\varphi,\mathcal{D}]\rvert_{\textrm{on-shell}} = \int_{\mathcal{D}\cap\partial\mathcal{M}} \mathcal{P}[\varphi,\partial_\mu\varphi]\,\delta\varphi$, where $\mathcal{P}$ is the canonical momentum density, a 3-form, of 
$\varphi$. 
Hence, stationarity 
also
implies the boundary condition $\iota^*\mathcal{P} = 0$, where $\iota:\partial\mathcal{M}\to\mathcal{M}$ is the inclusion map.

An instructive example is the relativistic open string \cite{Polchinski:1998rq} whose action is $\mathcal{S}[X,\mathcal{M}]=-\frac{1}{2}\int_{\mathcal{M}} d\tau d\sigma\eta^{ab}\partial_a X^\mu \partial_\beta X_\mu$ in conformally flat gauge, where $\mathcal{M}$ is the string worldsheet, \textit{i.e.}, a compact surface with a boundary $\partial\mathcal{M}$ at $\sigma=0,\pi$. Varying the action about any solution of the bulk equations of motion and integrating by parts, we find a boundary contribution $\delta\mathcal{S}[X,\mathcal{M}] = -\int_{\partial\mathcal{M}} d\tau\,\delta X_\mu\partial_\sigma X^\mu$. Requiring that this vanish, for arbitrary $\delta X_\mu$, imposes that $\partial_\sigma X^\mu=0$ at the string endpoints. Together with the conformally flat gauge condition (that the induced worldsheet metric is $\propto \eta_{ab}$, hence $\partial_\tau X^\mu\partial_\tau X_\mu +\partial_\sigma X^\mu\partial_\sigma X_\mu = 0$), we infer that the trajectory of the string endpoint is null. 

\section{The action principle for gravity}

Let us now extend this action principle to gravity coupled to matter.  To gain some mathematical perspective, it is helpful to consider the gravitational action, including boundary terms, in general dimension $D=n+1$. We write the metric as $\bm{g}=\eta_{ab}\bm{e}^{a}\otimes \bm{e}^{b}$ ($a,b=0,\ldots,n$), where $\eta_{ab}={\rm diag}(-1,1,\ldots,1)$ is the Minkowski metric, $\{\bm{e}^{a}\}$ is an orthonormal basis of 1-forms (the ``vielbein"), $\bm{\omega}^{a}_{\;\;b}$ is the corresponding Levi-Civita connection 1-form (the spin connection), and $\bm{\Omega}^{a}_{\;\;b}=d\bm{\omega}^{a}_{\;\;b}+\bm{\omega}^{a}_{\;\;c}\wedge\bm{\omega}^{c}_{\;\;b}=\frac{1}{2}R^{a}_{\;\;bcd}\bm{e}^{c}\wedge \bm{e}^{d}$ is its curvature.  It is convenient to also define the form $\bm{\mu}_{a_0\ldots a_{m}}=\frac{1}{(n-m)!}\bm{\mu}_{a_0\ldots a_{n}}\bm{e}^{a_{m+1}}\wedge\ldots\wedge \bm{e}^{a_{n}}$, where $\bm{\mu}_{a_0\ldots a_n}$ are the components of the volume form $\bm{\mu}$ relative to the basis $\{\bm{e}^a\}$. Assuming that this basis is positively oriented, then $\bm{\mu}_{a_0\ldots a_{n}}=\epsilon_{a_0\ldots a_{n}}$, where $\epsilon_{a_0\ldots a_{n}}$ is the totally anti-symmetric symbol with $\epsilon_{01\ldots n}=+1$. 

Lovelock \cite{Lovelock:1971yv,Lovelock:1972vz} found that the most general metric theory of gravity yielding second-order equations of motion is described the bulk action $\mathcal{S}[\mathcal{D}]=\int_{{\cal D}}\sum_{p=0}^{\floor*{D/2}}\alpha_{p}{\cal L}_{p}$ where ${\cal L}_{p}=\bm{\mu}_{a_{1}b_{1}\ldots a_{p}b_{p}}\bm{\Omega}^{a_1 b_1} \wedge \ldots \wedge \bm{\Omega}^{a_p b_p}$ is the $p$th Euler density, and $\alpha_{p}$ is an arbitrary real coefficient. Given that any open region with compact closure ${\cal D}$ has a boundary $\partial{\cal D}$, we should add to this the naturally corresponding boundary action $\mathcal{S}[{\partial{\cal D}}]=-\int_{\partial{\cal D}}\sum_{p=1}^{\floor*{D/2}}\alpha_p{\cal Q}_{p}$, where ${\cal Q}_{p}$ is the Chern-Simons form corresponding to the Euler density ${\cal L}_{p}$ (so, for any $2m$-dimensional compact manifold ${\cal M}$ with boundary $\partial{\cal M}$, the Euler characteristic is $\chi(\mathcal{M})=\int_{{\cal M}}{\cal L}_m-\int_{\partial{\cal M}}{\cal Q}_m$, by the Chern-Gauss-Bonnet theorem \cite{chern:1944gaussbonnet, chern1945curvatura, eguchi1980gravitation}; see \cite{myers1987higher} for helpful explicit formulae).

\begin{figure}%[ht!]
    \centering
    \includegraphics[width=0.45\linewidth]{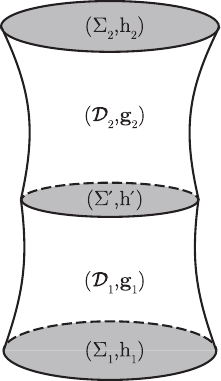}
    \caption{The gluing of two regions $({\cal D}_{1},\mathbf{g}_1)$ and $({\cal D}_{2},\mathbf{g}_2)$ at a common spacelike boundary $(\Sigma',h')$.}
    \label{fig:composition-rule}
\end{figure}

An important feature of the boundary term is emphasized by Hawking \cite{Hawking:1980gf}. Consider gluing two open regions ${\cal D}_1$ and ${\cal D}_2$ together at a common boundary $\Sigma'$ (see Fig.~2). For the metric to be continuous, the metrics on $({\cal D}_1,\mathbf{g}_1)$ and $({\cal D}_2,\mathbf{g}_2)$ must induce the same metric $h'$ on $\Sigma'$. It is then natural to require that the action is additive, \textit{i.e.}, the action for ${\cal D}_1\cup{\cal D}_2$ should be the sum of the actions for ${\cal D}_1$ and ${\cal D}_2$.  In the path integral formulation of quantum gravity, the amplitude to go from spatial slice $\Sigma_1$ (with metric $h_1$) to spatial slice $\Sigma_2$ (with metric $h_2$) is given by the path integral of ${\rm e}^{i{\cal S}}$ over all intervening space-times $({\cal M},\mathbf{g})$ starting at $(\Sigma_1,h_1)$ and ending at $(\Sigma_2,h_2)$. The additivity of the action ensures that path integral amplitudes satisfy the usual quantum mechanical composition rule 
\begin{equation}
    \langle \Sigma_2,h_2|\Sigma_1,h_1\rangle = \sum_{(\Sigma',h')} \langle \Sigma_2,h_2|\Sigma',h'\rangle \langle \Sigma',h'|\Sigma_1,h_1\rangle,
\end{equation}
when we sum over all intermediate 3-geometries $(\Sigma',h')$.  

The bulk action ${\cal S}[{{\cal D}}]$ does {\it not} satisfy this condition. However, including the boundary term by defining ${\cal S}_{\textrm{gr}}[\mathcal{D}]={\cal S}[{{\cal D}}]+{\cal S}[{\partial{\cal D}}]$ precisely fixes the problem.  Thus, this combination is the most natural gravitational action, and the one we shall consider here.  Specializing to 4D space-time, the Lagrangian consists of three bulk terms: ${\cal L}_0$, ${\cal L}_1$ and ${\cal L}_2$ (the cosmological constant, Einstein-Hilbert, and Gauss-Bonnet terms, respectively); and two boundary terms: ${\cal Q}_1$ and ${\cal Q}_2$ (the Gibbons-Hawking-York term \cite{york1972boundary,Gibbons:1976ue} and Myers term \cite{myers1987higher}, respectively). Since both the Gauss-Bonnet term $\int{\cal L}_2$ and its counterpart ${\cal Q}_2$ are topological, they may be important quantum mechanically, but may be ignored when considering the classical variational problem. Hence, the classically allowed configurations are governed by
\begin{align}
    \mathcal{S}_{\textrm{gr}}[\mathbf{g},\mathcal{D}] := \mathcal{S}_{\textrm{EH}}[\mathbf{g},\mathcal{D}] + \underbrace{\frac{\epsilon}{8\pi}
    \int_{\partial\mathcal{D}}\textrm{tr}_{\overline{g}}{K}\,\textrm{Vol}_{\overline{g}}}_{=\mathcal{S}_{\textrm{GHY}}[\mathbf{g},\mathcal{D}]},
    \label{eqn:grav-eq}
\end{align}
where $\mathcal{S}_{\textrm{EH}}:=\int\frac{1}{16\pi}\star(R-2\Lambda)$ is the Einstein-Hilbert action, $\mathcal{S}_{\textrm{GHY}}$ is the Gibbons-Hawking-York action, $\overline{g}$ is the induced metric on $\partial\mathcal{D}$, $n$ is the outward-pointing unit normal,  $K := \frac{1}{2}\pounds_n\overline{g}$ is the second fundamental form of $\partial\mathcal{D}$ and $\epsilon = g(n,n)=+1$ ($-1$) for spacelike (timelike) $n$. 

In order to properly extend \eqref{eqn:grav-eq} to a singular boundary, one must keep in mind that we cannot vary the action with respect to the metric $\mathbf{g}$ at the singular boundary, as it may be ill-defined there. To overcome this difficulty, we use the \textit{tetrad formalism} \cite{Straumann:2013spu,Krasnov_2020}.  To check whether a metric $\mathbf{g}$ that solves the bulk equations of motion {\it also} satisfies the requisite boundary condition at its spacelike singularity, we first express $\mathbf{g}$ in the vicinity of its singular boundary in a synchronous-like gauge:
\begin{equation}
    \mathbf{g} = -dt\otimes dt + g_{ij}\,e^i\otimes e^j,
\end{equation}
where $t\geq0$, with $t=0$ at the 
singularity and  $e^i$, $i=1,2,3$ are 1-forms that, together with $dt$, form a co-frame in any neighborhood of the singularity. The 3-metric $g_{ij}(t,x)$ is a fixed classical solution of the bulk Einstein-matter equations in synchronous gauge. We then consider variations of the coframe $e^i$, keeping $dt$ \textrm{and} $g_{ij}$ fixed. These variations have the form $\delta e^i = {f^i}_j\,e^j$, with ${f^i}_j$ arbitrarily chosen, smooth compactly supported functions, and induce corresponding metric variations $\delta \mathbf{g} = 2f_{(ij)} e^i\otimes e^j$. 

Provided that the variation of the Lagrangian of the matter fields does not yield boundary terms (which is the case, for example, for perfect fluids), the on-shell variation of the total action is given by: 
\begin{align}
    \label{eqn:var-term-cond}
    \delta_{\mathbf{e}}\mathcal{S}[\mathbf{g},\mathcal{D}]  & = \frac{1}{16\pi}\lim_{t\downarrow 0^+}\int_{\mathcal{D}\cap\Sigma_t} 
    ({K^i}_j - \textrm{tr}_{\overline{g}}K{\delta^i}_j){f^j}_i\,\overline{\mu}_{t},
\end{align}
where $\delta_{\mathbf{e}}$ represents the variation with respect to the co-frame $\{e^i\}$, and $\overline{\mu}_t$ is the induced volume form of the constant time slice $\Sigma_t$. 
Since the matrix-valued variation ${f^j}_i$ is arbitrary and smooth up to the boundary, \eqref{eqn:var-term-cond} is equivalent to $\left({K^i}_j-\textrm{tr}_{\overline{g}}K\,{\delta^i}_j\right)\overline\mu_t\to0$ in the limit $t\to 0$. Taking the trace of this equation yields $\textrm{tr}_{\overline{g}}\,K\,\overline{\mu}_t\to 0$ in the limit $t\to 0$, which then gives us the condition 
\begin{equation}
    \label{eqn:cosmology-cond}
    {K^i}_j\overline{\mu}_t\to 0\quad {\rm as}\quad  t \downarrow 0^+.
\end{equation}
Note that one is not allowed to lower or raise indices in \eqref{eqn:cosmology-cond}, if the metric tensor is ill-defined at the singularity.

\section{Application to cosmology}

Let us now apply this formalism to some interesting cosmological solutions of the Einstein-fluid field equations. One class of space-times that {\it do} satisfy \eqref{eqn:cosmology-cond} are radiation-dominated cosmologies with $P=\frac{1}{3}\rho$, emerging from isotropic singularities with line elements of the form  
\begin{equation}
    \label{eqn:conf-reg-newman}
    \mathbf{g} = -dt\otimes dt + a^2(t)\gamma_{ij}(t,x) \,dx^i\otimes dx^j,
\end{equation}
with $a(t)=t^{\frac{1}{2}}$, $\gamma_{ij}(t,x)$ a Riemannian 3-metric regular at $t=0$, ${K^i}_j= (1/2t){\delta^i}_j+\mathcal{O}(t)$, $\overline{\mu}_t= t^{3/2}\textrm{Vol}_\gamma$, where $\textrm{Vol}_\gamma$ is the volume form of the 3-metric $\gamma$, and therefore 
\begin{equation}
    {K^i}_j\,\overline{\mu}_t = 
    \frac{1}{2}t^{1/2} \textrm{Vol}_\gamma + \mathcal{O}(t^{3/2})\rightarrow 0, \quad {\rm as}\quad  t \downarrow 0^+.
\end{equation}
These are the \textit{conformally-regular geometries} studied by Tod, Newman and others~\cite{todd1991isotropic0,newman1993conformal1,newman1993conformal2}.

A family of geometries that do \textit{not} satisfy \eqref{eqn:cosmology-cond} consists of space-times whose induced volume form on constant $t$ slices, in the \textit{Constant Mean Curvature (CMC)} gauge, satisfies $\overline{\mu}_t = \mathcal{O}(t^\delta)$, where  $0<\delta\leq 1$. Since $\textrm{tr}_{\overline{g}}K \sim 1/t$, it follows that $\textrm{tr}_{\overline{g}}K\,\overline{\mu}_t = \mathcal{O}(t^{\delta-1})\neq 0$ in the limit $t\to 0$. A subclass of such geometries are homogeneous (but anisotropic) \textit{Kasner-like geometries} of the form 
\begin{equation}
    \label{eqn:geom-kasner}
    \mathbf{g} = -dt\otimes dt + \sum_{i=1}^3 t^{2p_i}e^i\otimes e^i,
\end{equation}
where $\Sigma$ is a 3-manifold with co-frame $e^i$, and the $p_i$ are functions on $\Sigma$ satisfying $\sum_i p_i = 1$. Solutions like these occur both in pure gravity (where $\sum_i p_i^2 = 1$) and also for stiff matter, or massless scalars, where $\sum_i p_i^2 < 1$ \cite{Andersson:2000cv,Fournodavlos:2020jvg,Fournodavlos:2020tti,Athanasiou:2024pzg}.  A much more general class of such geometries are the so-called \textit{BKL spacetimes} \cite{Belinsky:1970ew, Belinsky:1982pk,2001PhLB..509..323D,Perry:2023hiq}, which are conjectured to represent the generic behavior of a metric near a spacelike singularity: since these geometries are locally Kasner-like near the singularity, they are also incompatible with \eqref{eqn:cosmology-cond}.

The next result (proven in the Appendix) shows that, for nearly isotropic FLRW backgrounds and linear perturbations, \eqref{eqn:cosmology-cond} is equivalent to conformal regularity at the bang together with reflecting boundary conditions for the admissible perturbations.

\vspace{0.5em}

\noindent
\textbf{Theorem.} Suppose $(\Sigma,\gamma)$ is a closed, homogeneous and isotropic 3-manifold without boundary and constant curvature $\mathcal{K}=0,\pm 1$, and $(\mathcal{M},{}^0\mathbf{g})$ is the corresponding FLRW solution to the Einstein field equations, where the matter is modeled by a perfect fluid with equation of state $P=w\rho$, with $w$ a constant.

\begin{enumerate}
\item If $0\leq w<1$, the solution $(\mathcal{M},{}^0\mathbf{g})$ satisfies \eqref{eqn:cosmology-cond}, and is therefore a critical point of the action. For stiff fluids, $w=1$, the on-shell action is not finite.

\item For $0\leq w<1$, the perturbations $\delta\mathbf{g}$ that satisfy the linearized version of \eqref{eqn:cosmology-cond} (Eq. \eqref{eqn:cosm-var-cond} below) obey
\begin{equation}
    \label{eqn:pt-symmetry}
    \frac{\partial \phi_B}{\partial\eta}\bigg\rvert_{\eta = 0} = \frac{\partial h_{ij}}{\partial\eta}\bigg\rvert_{\eta = 0} = 0,
\end{equation}
where $\eta$ is the conformal time, $\phi_B$ is the Bardeen potential, and $h_{ij}$ are the tensor perturbations. No vector perturbations are allowed and the perturbed geometry is conformally regular.  In the special cases of dust or perfect radiation, the perturbations are analytic in $\eta$ at $\eta=0$ and invariant under the reflection $\eta\to -\eta$.
\end{enumerate}

\section{Discussion}

Motivated by the path integral formulation of quantum gravity, we have proposed that the action principle for general relativity be extended to space-times whose boundary includes a spacelike singularity. This extension leads to the on-shell boundary conditions \eqref{eqn:cosmology-cond} at a singular boundary, which restrict the admissible initial data. In particular, these conditions favor conformally regular geometries (like FLRW hot big bang models) and exclude anisotropic Kasner-like and chaotic BKL-type behavior. For linear perturbations around FLRW cosmologies, we have shown that the same variational principle is equivalent to imposing reflecting boundary conditions at the big bang. Newman and others have studied the nonlinear Einstein-fluid equations for a radiation-dominated universe near 
conformally-regular 
singularities where the freely specifiable data reduce to the initial conformal 3-geometry~\cite{todd1991isotropic0,newman1993conformal1,newman1993conformal2,newman1998sing,anguige1999isotropic1,anguige1999isotropic2}, with similar findings. This leads us to conjecture that the variational condition \eqref{eqn:cosmology-cond} selects the conformally regular 4-geometries more generally.

The action principle presented here provides a variational 
underpinning
for $CPT$-symmetric cosmologies \cite{boyle2018cpt,boyle2018bigbang,Turok:2022fgq, Boyle:2022lcq}. Instead of enforcing  $CPT$-symmetric boundary conditions at the big bang via the method of images, applied across the bang, the same $CPT$-symmetric reflecting boundary conditions here emerge 
from
the gravitational action principle when we treat the big bang as a free boundary. With this boundary condition, a $CPT$-symmetric universe naturally appears as the dominant semi-classical saddle of the Lorentzian path integral. 

We are thus led to explore an alternative to Hartle and Hawking's no-boundary proposal \cite{Hartle:1983ai}, namely the conjecture that the initial boundary to space-time was free. In both cases, initial conditions at the big bang are not imposed by hand, but are derived from a simple postulate concerning the gravitational path integral. The implementation of the two proposals are, however, quite different. In the no-boundary proposal, the wavefunction of the universe for a prescribed spatial closed 3-geometry $(\Sigma,h)$ is defined by a formal integration over a class of compact \textit{Euclidean} 4-geometries whose only boundary is $(\Sigma,h)$. The corresponding semi-classical boundary conditions for the geometry and matter fields are then obtained from this Euclidean prescription. Here, by contrast, we begin with the \textit{Lorentzian} gravitational path integral in the semi-classical approximation and ask whether the saddle-point approximation can remain meaningful when a singular boundary is included. We find that it can, provided that the singularity satisfies the \textit{free boundary condition} \eqref{eqn:cosmology-cond}. The admissible saddles are then precisely the conformally regular geometries with reflecting initial conditions.

By being formulated in Lorentzian signature, our proposal avoids some of the technical difficulties that the no-boundary proposal faces~\cite{Feldbrugge:2017kzv,Feldbrugge:2017fcc,Feldbrugge:2017mbc,Lehners:2023yrj}. It falls short of being a fully quantum prescription for the initial boundary of the universe but does, perhaps, suggest a natural route towards such a prescription. In canonical quantum gravity, where the 3-geometry $\overline{g}$ is the configuration variable, its conjugate momentum is given by $\Pi^{ij} := (K^{ij}-\textrm{tr}_{\overline{g}}K\,\overline{g}^{ij})\overline{\mu}.$ 
Equivalently, lowering one index with $\overline{g}$, the relevant mixed momentum is ${\Pi^i}_j$. To generalize \eqref{eqn:cosmology-cond} to quantum cosmology, one would have to resolve the operator ordering problem in both the boundary condition and the constraints imposed by diffeomorphism invariance.

Note that the boundary condition (\ref{eqn:cosmology-cond}) does {\it not} apply to boundaries
at infinite (spacelike or timelike) geodesic distance (such as the AdS boundary, or the asymptotically dS boundary of our own universe), since a compact region ${\cal D}$ of a (pseudo-)Riemannian manifold has finite volume, and hence cannot also intersect a boundary at infinity.  The fact that Eq.~(\ref{eqn:cosmology-cond}) constrains the past (big bang) boundary of our universe to a special (low entropy) state, but does not constrain the future (asymptotically dS) boundary, neatly accords with the second law of thermodynamics.

However, the boundary condition (\ref{eqn:cosmology-cond}) \textit{does} apply to boundaries at finite (spacelike or timelike) geodesic distance, such as those formed inside a black hole by gravitational collapse. Given that the geometry of spherical collapse is Kasner-like, and generic collapse is conjectured to be of BKL type, these interior solutions are also incompatible with the free boundary condition \eqref{eqn:cosmology-cond}. 
On the other hand, the black mirror construction \cite{Tzanavaris:2024acr} in which two symmetric black hole exteriors are glued at their common boundary (which, in the stationary case is the event horizon), {\it is} compatible with \eqref{eqn:cosmology-cond} (since there is no interior).  In forthcoming work, we show that the generalization of the boundary condition \eqref{eqn:cosmology-cond} for null boundaries is satisfied at the black mirror's surface, so one can reinterpret the black-mirror as a one-sided solution with the black mirror's surface as an internal null boundary.  This generalizes the black mirror's surface in the fully dynamical (non-stationary) case, where it is no longer the same as the event horizon. 
Thus, the criterion \eqref{eqn:cosmology-cond} acts as a filter of sorts, not ruling out all singularities, but providing a selection criterion. This agrees with Horowitz and Myers' proposal \cite{Horowitz:1995ta} that 
quantum gravity should not eliminate all singularities, but rather restrict the class of ``physically acceptable'' singularities. 

\section{Acknowledgments}

We are grateful to Bruce Allen, Guillaume Bossard, Job Feldbrugge, Andr\'es Franco-Grisales, Badri Krishnan, Jean-Luc Lehners, and Hans Ringstr\"om for stimulating discussions that contributed significantly to this work.  LB and NT are supported by the STFC Consolidated Grant `Particle Physics at the Higgs Centre', and NT is supported by the Higgs Chair at the University of Edinburgh. Research at Perimeter Institute is supported by the Government of Canada, through Innovation, Science and Economic Development, Canada and the Province of Ontario through the Ministry of Research, Innovation and Science. 

\appendix 

\section{Proof of the equivalence between reflecting initial conditions and the free boundary condition}

Throughout, $(\mathcal{M},\mathbf{g})$ will denote a four-dimensional Lorentzian manifold, with the ``mostly positive'' signature convention; bold letters will denote tensor fields defined on $\mathcal{M}$, while non-bold letters will denote tensor fields defined on a hypersurface of $\mathcal{M}$; $\mathbf{e}_a=e_a^{\mu}\partial_{\mu}$ is a local frame (or vierbein) on the tangent bundle $T{\cal M}$;  
$\mathbf{e}^a=e^{a}_{\mu}dx^{\mu}$ is the corresponding dual co-frame on the cotangent bundle $T^{\ast}\!{\cal M}$ satisfying 
$\mathbf{e}^a(\mathbf{e}_b)=e^{a}_{\mu}e^{\mu}_{b}=\delta^{a}_{b}$; $\mathbf{D}$ is the Levi-Civita (metric-compatible, torsion-free) connection on $\mathcal{M}$, and $\bm{\omega}^{a}_{\;\;b}$ is the corresponding connection one-form.  We denote by $\bm{\mu}$ the volume 4-form of $\mathcal{M}$, and its interior product with elements of $\{\mathbf{e}_a\}$ by $\bm{\mu}_a = \bm{\mu}(\mathbf{e}_a,\cdot)$, $\bm{\mu}_{ab} = \bm{\mu}(\mathbf{e}_a,\mathbf{e}_b,\cdot)$, etc. Given any hypersurface, $\overline{\mu}$ will denote its induced volume form, and similarly, $\overline{\mu}_a :=\overline{\mu}(\mathbf{e}_a,\cdot),$ $\overline{\mu}_{ab} :=\overline{\mu}(\mathbf{e}_a,\mathbf{e}_b,\cdot),$ etc. Indices $i,j,...$ take values in $\{1,2,3\}$, while $a,b,...$ take values in $\{0,1,2,3\}$. Lastly, given any vector $\mathbf{u}=u^a \mathbf{e}_a$, we denote by $\mathbf{u}^\flat = u_a\mathbf{e}^a$ the 1-form resulting by ``lowering the index'' of $\mathbf{u}$.

To prove the theorem, we will briefly present an action principle for perfect fluids, as formulated in \cite{schutz1970fluid,tzanavaris:2026fluids}. The kinematic state of a one-component perfect fluid is determined by a timelike unit vector field $\mathbf{u}$, representing its velocity. Each worldline represents a fluid parcel. Thus, interpreting a perfect fluid as a timelike congruence means that its equations of motion are derived by varying each worldline individually. This is accomplished by locally labeling each worldline with \textit{Lagrangian coordinates} $\mathrm{a}^1,\mathrm{a}^2,\mathrm{a}^3$, that are by definition conserved along each flowline of the fluid, \textit{i.e.}, $\pounds_u\textrm{a}^i=0$ for all $i=1,2,3.$ 
%Taking into consideration 
%the laws of 
Considering conservation of mass and entropy, 
\begin{equation}
    \textrm{div}(n\mathbf{u}^\flat) = \textrm{div}(s\mathbf{u}^\flat) = 0,
\end{equation}
where $n$ and $s$ are the particle number and entropy density, 
%its 
the fluid's action is then given by
\begin{equation}
    \label{eqn:lag-dens-off-shell}
    \mathcal{L}_m = -\left[\star (\rho + \mathcal{A}) + \bm{\mathcal{C}}\wedge\star \mathbf{u}^\flat \right],
\end{equation}
where 
\begin{subequations}
\begin{align}
    \mathcal{A} & = \frac{1}{2}f\left(\mathbf{g}(\mathbf{u},\mathbf{u}) + 1\right), \label{eqn:affine-param}\\
    \bm{\mathcal{C}} & = n\,d\Phi + s\,d\Theta + b_i\,d\textrm{a}^i.
\end{align}    
\end{subequations}
Here, $\Phi,\Theta$ and $b_i$ are Lagrange multipliers. The variation with respect to the velocity $\mathbf{u}$ yields the Clebsch decomposition of the velocity 
\begin{equation}
    (\rho+P)\mathbf{u}^\flat = nd\Phi + sd\Theta + b_id\mathrm{a}^i,
\end{equation}
where $\rho$ and $P$ are the energy density and pressure, respectively. The variations with respect to $n$ and $s$ yield
\begin{equation}
    \mu = -\pounds_u\Phi, \quad T = -\pounds_u\Theta, 
    \label{eqn:therm-pot}\\
\end{equation}
where $\mu$ and $T$ are the chemical potential and temperature of the fluid. The on-shell action and its variation are given by 
\begin{subequations}
\begin{align}
    \mathcal{S}_m[\mathcal{D}] & = \int_{\mathcal{D}} P\textrm{Vol}_g,
    \\
    \delta\mathcal{S}_m[\mathcal{D}] & = \int_{\mathcal{D}} \mathbf{T}_{ab} 
    \delta \mathbf{e}^a\wedge\star \mathbf{e}^b
    \nonumber\\
    & \quad + \int_{\partial\mathcal{D}} (n\delta\Phi +  s\delta\Theta + b_i\delta\mathrm{a}^i)\star\mathbf{u}^\flat,
    \label{eqn:bound-term-fluid}
\end{align}    
\end{subequations}
where $\mathbf{T}=(\rho+P)\mathbf{u}^\flat\otimes\mathbf{u}^\flat + P\mathbf{g}$ is the stress-energy tensor of perfect fluids. Using the conservation laws, one can immediately deduce that the boundary term \eqref{eqn:bound-term-fluid} is finite at the big bang, and that $\Phi$ and $\Theta$ are the ``potentials'' of the thermodynamic variables $\mu$ and $T$, and we can always fix them at the big bang without over-constraining the physical variables. This leaves the boundary term $\int_{\partial\mathcal{D}}b_i\delta\mathrm{a}^i\star\mathbf{u}^\flat.$ This term can either vanish by fixing the initial positions of the fluid parcels $(\delta\mathrm{a}^i=0)$, or by setting $b_i=0$. For adiabatic fluids $(s=\textrm{const.})$, this implies that the vorticity of the fluid is zero. The vorticity is sourced by vector (metric) perturbations which, as we shall show later, are forced to vanish by the condition \eqref{eqn:cosmology-cond}, proving that the free boundary problem is self-consistent. 

Thus, to judge whether the solution is a critical point to the action, one has to determine whether the on-shell action is finite at any region $\mathcal{D}$. Taking the trace of Einstein's field equation and assuming non-zero cosmological constant, we have $R\sim \rho$. Using cosmic time, we have 
\begin{equation}
    a(t)\sim t^{2/3(1+w)}, \;\; 
    \rho(t)\sim t^{-2},\;\; \textrm{Vol}_{{}^0\mathbf{g}}\sim t^{2/(1+w)}\,dt\,d^3x,
\end{equation}
so that 
\begin{equation}
    \rho\textrm{Vol}_{{}^0\mathbf{g}}\sim t^{-2w/(1+w)}\,dt\,d^3x
\end{equation}
which is integrable for $0\leq w < 1$ and logarithmically divergent for $w=1$. 

In order to prove (2), we derive the first order correction to the boundary condition \eqref{eqn:cosmology-cond}. Given that,
\begin{align}
    {K^i}_j\overline{\mu}_t & = {}^0{K^i}_j {}^0\overline{\mu}_t + \delta({K^i}_j\overline{\mu}_t) + ...,
\end{align}
where the omitted terms involve higher order, it follows that the boundary condition for the perturbations is 
\begin{equation}
    \label{eqn:cosm-var-cond}
    \lim_{t\to 0^+} \delta({K^i}_j\overline{\mu}_t) = 0. 
\end{equation}
To correctly identify these perturbations, we will use the ADM ansatz
\begin{equation}
    \mathbf{g} = -N^2\,d\eta\otimes d\eta + g_{ij}(dx^i + S^i\,d\eta)\otimes(dx^j + S^j\,d\eta).
\end{equation}
The second fundamental form of the surfaces of constant $\eta$ is given by 
\begin{equation}
    K_{ij} = \frac{1}{2N}(\partial_\eta g_{ij} - 2\nabla_{(i}S_{j)}),
\end{equation}
where $\nabla$ is the induced connection, and the induced volume form is $\overline{\mu}_t = \sqrt{\det(g_{ij})}\,d^3x$. A straightforward but tedious calculation shows that for perturbations around FLRW, we have 
\begin{subequations}
\begin{align}
        \delta{K^i}_j & = \frac{1}{2a}g^{ik}\left(\partial_\eta\delta g_{kj} - 2\nabla_{(k}\delta S_{j)}\right)
        \nonumber\\
        \label{eqn:var-extr-cur}
        & \qquad\qquad\qquad - \frac{a'}{a^2}g^{ik}\,\delta g_{k\ell} - \frac{a'}{a^3}\delta N\,{\delta^i}_j,
        \\
        \delta\overline{\mu}_t & = \frac{1}{2}g^{k\ell}\,\delta g_{k\ell}\,\overline{\mu}_t = 
    \frac{1}{2}a^3 g^{k\ell}\,\delta g_{k\ell}\textrm{Vol}_{\gamma},
    \label{eqn:var-vol-form}
\end{align}
\end{subequations}
where $\gamma$ is a 3-metric of constant curvature $\mathcal{K}=0,\pm 1$ and $\textrm{Vol}_{\gamma} = \sqrt{\det(\gamma_{ij})}\, d^3x$ is the positively-oriented volume form corresponding to $\gamma$. 

\vspace{0.5em}

\noindent
\underline{Scalar perturbations.}

\vspace{0.5em}

\noindent
Because there is no anisotropic stress, the two Bardeen potentials $\phi_B,\psi_B$ are equal $(\phi_B=\psi_B=\phi)$, and the perturbations to the metric in the longitudinal gauge are 
\begin{equation}
    \delta\mathbf{g}_{00}^{(S)} = -2a^2\phi, \quad \delta\mathbf{g}_{ij}^{(S)} = -2a^2\phi\,\gamma_{ij},
    \label{eqn:scalar-perturb}
\end{equation}
which results in 
\begin{align}
    \label{eqn:scal-contr}
    \delta^{(S)}({K^i}_j\,\overline{\mu}_t) & = -\frac{1}{a^2}\partial_\eta(a^4\phi_B)\,{\delta^i}_j\textrm{Vol}_\gamma,
\end{align}
which is manifestly gauge-invariant. 

\vspace{0.5em}

\noindent
\underline{Vector perturbations.}

\vspace{0.5em}

\noindent
A general vector perturbation (see \cite{Stewart:1990fm}) is given by 
\begin{equation}
    \delta\mathbf{g}_{0i}^{(V)} = a^2B_i, \quad \delta\mathbf{g}_{ij}^{(V)} = 2\nabla_{(i}E_{j)},
\end{equation}
where $B,E$ are divergence free 1-forms. The gauge symmetry of vector perturbations is 
\begin{equation}
    B\to B+\partial_\eta\xi, \quad E\to E + \xi,
\end{equation}
where $\xi$ is an arbitrary, divergence-free 1-form with no time components. The gauge-invariant quantity is 
\begin{equation}
    \mathcal{V} : = B - \partial_\eta E. 
\end{equation}
Thus, the natural gauge choice is the \textit{vector gauge}, in which $E=0$ and $\mathcal{V} = B$. In this gauge, one can easily deduce that 
\begin{equation}
    \label{eqn:vec-contr}
    \delta^{(V)}({K^i}_j\overline{\mu}_t) = -a^2\nabla_{(i}\mathcal{V}_{j)}\,\textrm{Vol}_{\gamma}.
\end{equation}
It should be noted that $\mathcal{V}$ is unique up to a Killing vector, for if $\nabla_{(i}\mathcal{V}_{j)} = 0$, then under a gauge transformation from the vector gauge to the synchronous gauge: 
\begin{equation}
    B(\eta)\to 0, \quad E(\eta)\to \int^{\eta} ds\, \mathcal{V}(s,\cdot),
\end{equation}
the vector perturbations result in a zero contribution to the metric. 

\vspace{0.5em}

\noindent
\underline{Tensor perturbations.}

\vspace{0.5em}

\noindent
The tensor perturbations are automatically gauge invariant, and are given by 
\begin{equation}
    \delta\mathbf{g}_{ij}^{(T)} = a^2h_{ij},
\end{equation}
where $h_{ij}$ is transverse and traceless, \textit{i.e.}, 
\begin{equation}
    {h^i}_i = \nabla_i {h^i}_j = 0. 
\end{equation}
Then, a straightforward calculation shows 
\begin{equation}
    \label{eqn:tens-contr}
    \delta^{(T)}({K^i}_j\overline{\mu}_t) = \frac{1}{2}a^2\partial_\eta {h^i}_j\,\textrm{Vol}_\gamma
\end{equation}

Having expressed the perturbation $\delta({K^i}_j\overline{\mu}_t)$ in gauge-invariant form, the next step is to investigate the implication of the conditions $\delta({K^i}_j\overline{\mu}_t)\to 0$ as $t\to 0$. We will show that these conditions imply $\mathcal{V} = 0$, and that $\phi_B$ and $h_{ij}$ satisfy \eqref{eqn:pt-symmetry}. The proof is as follows. First, we show that the contribution \eqref{eqn:scal-contr} of the singular scalar modes diverge, whereas the contributions \eqref{eqn:vec-contr} and \eqref{eqn:tens-contr} of singular vector and tensor modes are bounded. This discards the singular scalar modes. The last step in proving \eqref{eqn:pt-symmetry} is to show that the vector and tensor contributions cannot cancel each other out, which would imply that the singular modes must vanish individually. The regular modes, \textit{i.e.}, those that are bounded at $\eta = 0$, are precisely those that satisfy \eqref{eqn:pt-symmetry} (see \cite{boyle2018cpt,boyle2018bigbang}). To complete the form, we will extract a perturbation in the synchronous gauge, in which the condition \eqref{eqn:cosmology-cond} was formulated. 

\vspace{0.5em}

\noindent
\underline{Step 1. $\delta^{(S)}({K^i}_j\overline{\mu}_t)$ diverges.} 

\vspace{0.5em}

\noindent
Assuming that the perturbations are adiabatic, the equation for the Bardeen potential is of the form 
\begin{equation}
    \left[\partial_\eta^2 + 3(1+w)\mathcal{H}\partial_\eta + \hat{F}\right]\phi_B  = 0,
\end{equation}
where $\hat{F} = -\Delta_\gamma + f$, and $f$ is a function of $\eta$ that is continuous at $\eta=0$. In the limit $\eta\to 0$, we have, 
\begin{equation}
    a(\eta)\sim \eta^{2/(1+3w)},\quad\mathcal{H}\sim\frac{2}{(1+3w)\eta},
\end{equation}
and therefore the asymptotic behaviour of $\phi_B$ is described by the equation 
\begin{equation}
    \left(\partial_\eta^2 + \frac{6(1+w)}{1+3w}\frac{1}{\eta}\partial_\eta \right)\phi_B = 0. 
\end{equation}
The asymptotics for the derivative, and consequently $\delta^{(S)}({K^i}_j\overline{\mu}_t)$, are derived by integrating the equation above. The ones whose derivative is non-zero at $\eta=0$ scale as 
\begin{equation}
    \partial_\eta\phi_B\sim \eta^{-6(1+w)/(1+3w)},\quad 
    \delta^{(S)}({K^i}_j\overline{\mu}_t)\sim \eta^{-2},
\end{equation}
thus showing that their contribution to the boundary condition indeed diverges.

\vspace{0.5em}

\noindent
\underline{Step 2. $\delta^{(V)}({K^i}_j\overline{\mu}_t)$ is bounded.}

\vspace{0.5em}

\noindent
The vector gauge-invariant vector perturbation is obtained by integrating the equation 
\begin{equation}
    \partial_\eta \nabla_{(i}\mathcal{V}_{j)} + 2\mathcal{H} \nabla_{(i}\mathcal{V}_{j)} = 0
\end{equation}
which is given by (A2.9) in \cite{Stewart:1990fm}, and the fact that there is no anisotropic stress, and gives us 
\begin{equation}
    \nabla_{(i}\mathcal{V}_{j)} = \frac{1}{a^2}F_{ij},
\end{equation}
where $F_{ij}$ is a symmetric tensor satisfying $\partial_\eta F_{ij}=0$. Simple substitution shows 
\begin{equation}
    \delta^{(V)}({K^i}_j\overline{\mu}_t) = -F_{ij}\,\textrm{Vol}_{\gamma},
\end{equation}
and therefore its contribution to the boundary action is constant. 

\vspace{0.5em}

\noindent
\underline{Step 3. $\delta^{(T)}({K^i}_j\overline{\mu}_t)$ is bounded.}

\vspace{0.5em}

\noindent
The equation describing the tensor modes is given by 
\begin{equation}
    \left(\partial_\eta^2 + 2\mathcal{H}\partial_\eta + \hat{\mathcal{F}}\right) h_{ij} = 0,
\end{equation}
where $\hat{\mathcal{F}} = -\Delta_\gamma + 2\mathcal{K}$. As with the scalar modes, the asymptotic behaviour of the tensor modes is described by the equation
\begin{equation}
    \left(\partial_\eta^2 + \frac{4}{1+3w}\frac{1}{\eta}\partial_\eta\right)h_{ij} = 0.
\end{equation}
Integrating this equation yields 
\begin{equation}
    \partial_\eta h_{ij} \sim \eta^{-4/(1+3w)}\sim a^{-2},
\end{equation}
proving that $\delta^{(T)}({K^i}_j\overline{\mu}_t)$ is indeed bounded at $\eta\to 0$. 

\vspace{0.5em}

\noindent
\underline{Step 4. Elimination of all singular modes.}

\vspace{0.5em}

\noindent
Suppose $\delta^{(V)}({K^i}_j\overline{\mu}_t)=-\delta^{(T)}({K^i}_j\overline{\mu}_t)$. Then, it follows that the $F_{ij}$ is transverse-traceless, as the limit of a transverse-traceless tensor. Taking the contraction of the Riemann tensor for $\nabla$, we have 
\begin{equation}
    \nabla^i\nabla_{(i}\mathcal{V}_{j)} = \frac{1}{2}(\Delta_\gamma + 2\mathcal{K})\mathcal{V}_j = 0,
\end{equation}
where $\mathcal{K}$ is the (constant) curvature of $(\Sigma,\gamma)$. On the other hand, since $\Sigma$ is a closed manifold without boundary, the divergence theorem implies 
\begin{align}
    \int_{\Sigma}\nabla_{(i}\mathcal{V}_{j)} & \nabla^{(i}\mathcal{V}^{j)}\,\textrm{Vol}_\gamma = 
    \nonumber\\
    &=-\frac{1}{2}\int_{\Sigma} \mathcal{V}^i\left(\Delta_\gamma+2\mathcal{K}\right)\mathcal{V}_i\,\textrm{Vol}_\gamma = 0.
\end{align}
This in turn leads to, $\nabla_{(i}\mathcal{V}_{j)} = 0$, and therefore $F_{ij} = 0$. Since vector modes cannot cancel the nonzero boundary contributions of singular tensor modes, all singular modes must vanish identically

\vspace{0.5em}

\noindent
\underline{Step 5. Derivation of the explicit perturbation.}

\vspace{0.5em}

\noindent
Without loss of generality, we can assume that the co-frame is the coordinate co-frame, \textit{i.e.}, $e^i=dx^i$. The general, non-singular perturbations to $\{e^i\}$ are given by 
\begin{equation}
    \delta e^i = \left(\nabla^i\nabla_jE - \phi{\delta^i}_j  + 
    \frac{1}{2}{h^i}_j \right)\,dx^j,
\end{equation}
where the indices raised by $\gamma_{ij}$, and
\begin{subequations}
\begin{align}
    \phi(\eta,\cdot) & = \phi_B(\eta,\cdot) + \frac{\mathcal{H}(\eta)}{a(\eta)}\int_0^\eta ds\, a(s)\phi_B(s,\cdot), \\
    E(\eta,\cdot) & = -\int_0^{\eta}\frac{ds}{a(s)}\int_0^{s} d\tau\,a(\tau)\phi_B(\tau,\cdot),
\end{align}    
\end{subequations}
To conclude the proof, we need to show that $\phi$ and $E$ are finite. For sufficiently small $\eta_0>0$, we have
\begin{subequations}
\begin{align}
    & |\phi(\eta)|\lesssim \sup_{0\leq\eta\leq\eta_0} |\phi_B(\eta)|+\eta\mathcal{H}(\eta)<+\infty, \\
    \nonumber\\
    & |\nabla^i\nabla_j\nabla E(\eta,\cdot)|\lesssim 
    \sup_{0\leq\eta\leq\eta_0}\eta^2|\nabla^i\nabla_j \phi_B(\eta)| < +\infty,
\end{align}    
\end{subequations}
proving that the variations $\{\delta e^i\}$ are well-defined, assuming sufficiently regular initial data at $\eta=\eta_0$.

\bibliographystyle{apsrev4-2}
\bibliography{bibliography}

\end{document}